\documentclass[11pt,twoside]{article}
\usepackage{cspm-asp2006}
\usepackage{epsfig,graphicx,natbib,url}  
\usepackage{lscape}
\begin{document}

\markboth{V.H. Hansteen et al.}{3d Models of the Solar Atmosphere}
\title{
3d Numerical Models of the Chromosphere, Transition Region, and Corona}
\author{Viggo H. Hansteen, 
        Mats Carlsson,
        Boris Gudiksen}
\affil{Institute of Theoretical Astrophysics, University of Oslo, Norway}

\begin{abstract} 
A major goal in solar physics has during the last five decades been to
find how energy flux generated in the solar convection zone is
transported and dissipated in the outer solar layers. Progress in this
field has been slow and painstaking. However, advances in computer
hardware and numerical methods, vastly increased observational
capabilities and growing physical insight seem finally to be leading
towards understanding. Here we present exploratory numerical MHD models
that span the entire solar atmosphere from the upper convection zone to
the lower corona. These models include non-grey, non-LTE radiative
transport in the photosphere and chromosphere, optically thin radiative
losses as well as magnetic field-aligned heat conduction in the
transition region and corona.
\end{abstract}

\section{Introduction}

The notion that chromospheric and coronal heating in some way follow 
from excess ``mechanical'' energy flux as a result of convective motions
has been clear since the mid-1940's. Even so, it is only recently that 
computer power and algorithmic developments have allowed one to even consider 
taking on the daunting task of modeling the entire system from convection
zone to corona in a single model. 

Several of these challenges were met during the last few years in 
the work of Gudiksen \& Nordlund (\cite{Gudiksen+Nordlund2002}), where it was 
shown that it is possible to model the photosphere to corona system. 
In their model a scaled down longitudinal magnetic field taken from an 
SOHO/MDI magnetogram of
an active region is used to produce a potential magnetic field in the
computational domain that covers 50$\times$50$\times$30~Mm$^3$. This
magnetic field is subjected to a parameterization of horizontal
photospheric flow based on observations and, at smaller scales, on
numerical convection simulations as a driver at the lower boundary.

In this paper we will consider a similar model, but one which simulates a smaller 
region of the Sun; at a higher resolution and in which convection is included. The 
smaller geometrical region implies that several coronal phenomena cannot be modeled.
On the other hand the greater resolution and the inclusion of convection (and the 
associated non-grey radiative transfer) should allow the model described here to 
give a somewhat more satisfactory description of the chromosphere and, perhaps, the
transition region and lower corona. 

\begin{figure}  
  \centering
  \includegraphics[width=\textwidth]{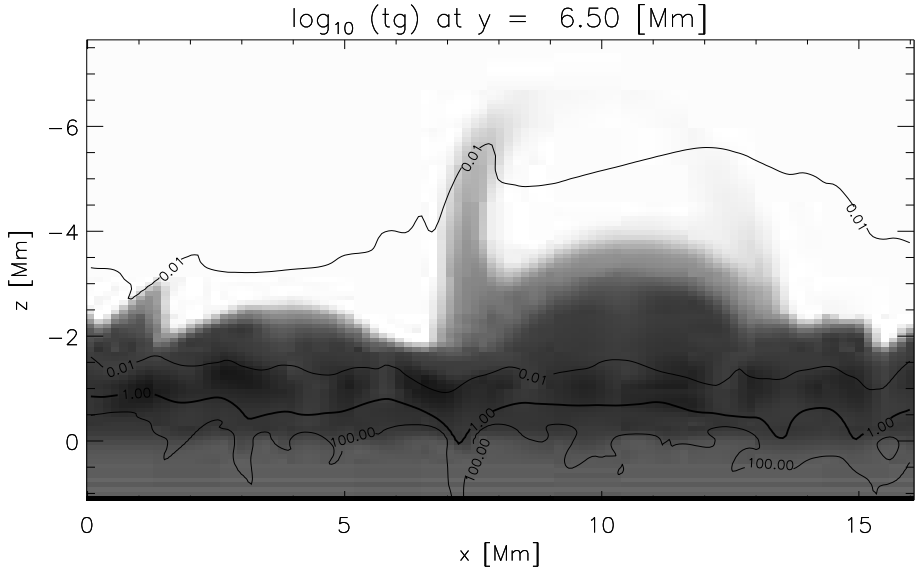}
  \caption[]{\label{hansteen-fig:qsmag-256t_xz_tg}
    Temperature structure in one plane of the simulation box.
    At $z=0$~Mm temperatures span 5000~K to $10\,000$~K, at the bottom
    of the computational domain, $1.5$~Mm into the convection zone, we find 
    temperatures ranging from $15\,700$~K in down-flowing plumes to $16\,500$ in 
    the gas flowing into the simulation domain. Above the photosphere, 
    the chromosphere extends 2--4~Mm (see also Figure~\ref{hansteen-fig:tdist-chrom})
    with temperatures from 2000~K to 8000~K.
    The upper 10~Mm of the model is filled with plasma at transition region and coronal 
    temperatures of up to some 1~MK, though even at great heights there are 
    intrusions of cooler gas. Also plotted are contours
    of plasma $\beta=p_g/p_B$; $\beta=1$ is found in the mid to lower chromosphere, 
    occasionally almost reaching the photosphere. $\beta$ is much greater than one
    in the convection zone, much smaller than one in the transtion region and corona.
    Note that we find high intensity ``bright points'' in the photosphere where the
    magnetic field is strong. High intensity in the chromosphere is due both 
    hydrodynamic shocks and magnetic heating.
}\end{figure}

\begin{figure}  
  \centering
  \includegraphics[width=\textwidth]{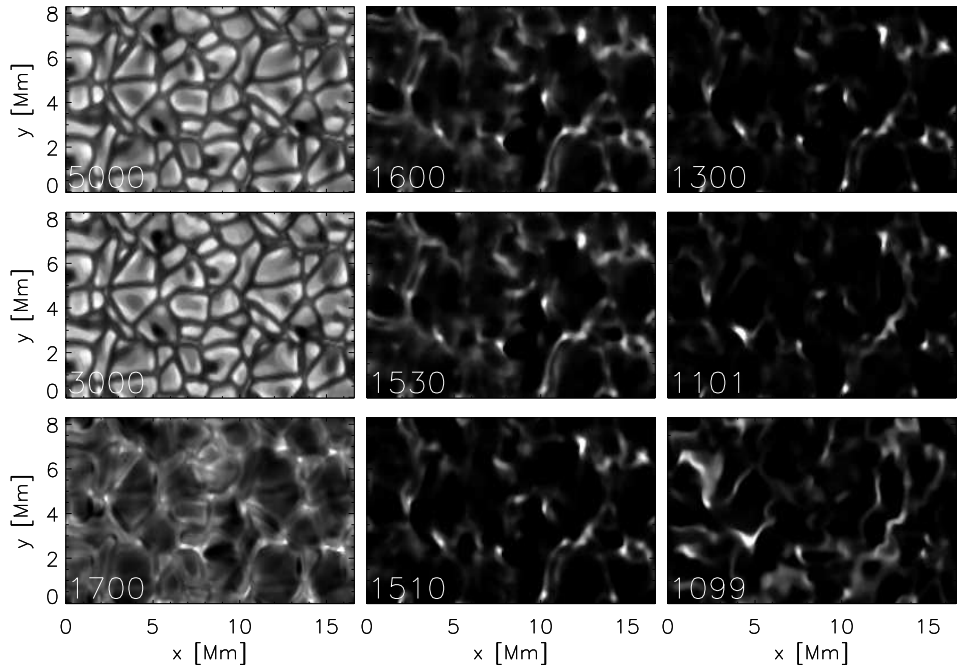}
  \caption[]{\label{hansteen-fig:fig_icont}
    Emergent intensities in various continua (wavelength in {\AA} in each panel)
    as seen from above. The continua at 5000 and 3000~{\AA}
    show photospheric granulation, the continuum at 1700 is formed
    a few hunderd kilometers higher and shows reverse granulation,
    at 1600 and 1530 {\AA} we sample the 
    the upper photosphere. The 1510 continuum is just shortward 
    of the bound-free edge of neutral silicon, carrying the formation
    to the lower chromosphere. The 1300 and 1101 continua are formed
    in the lower-mid chromosphere and the 1099 continuum is just shortward of
    the bound-free edge of neutral carbon and is formed in the
    upper chromosphere.
}\end{figure}

\begin{figure}  
  \centering
  \includegraphics[width=\textwidth]{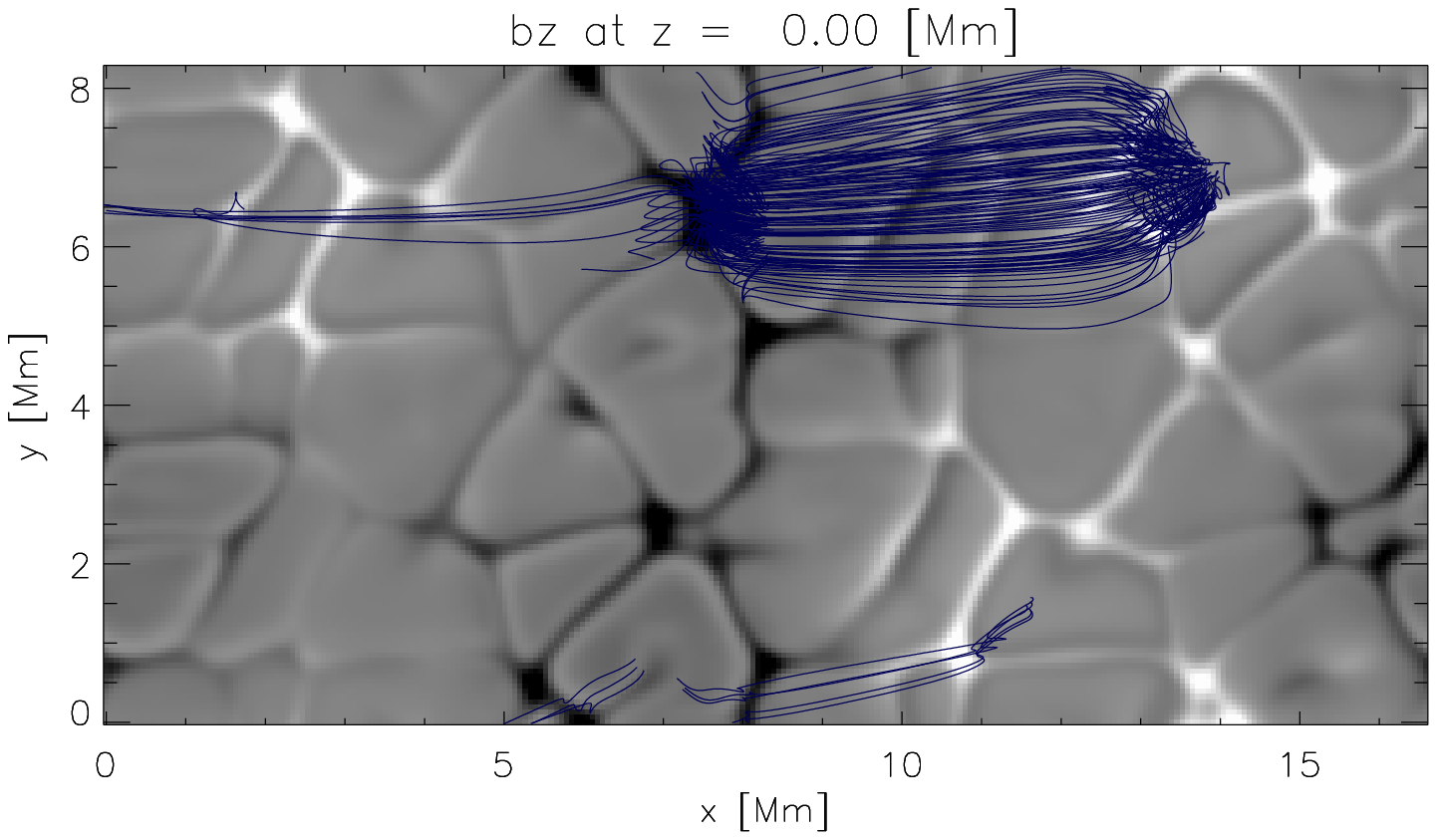}
  \includegraphics[width=\textwidth]{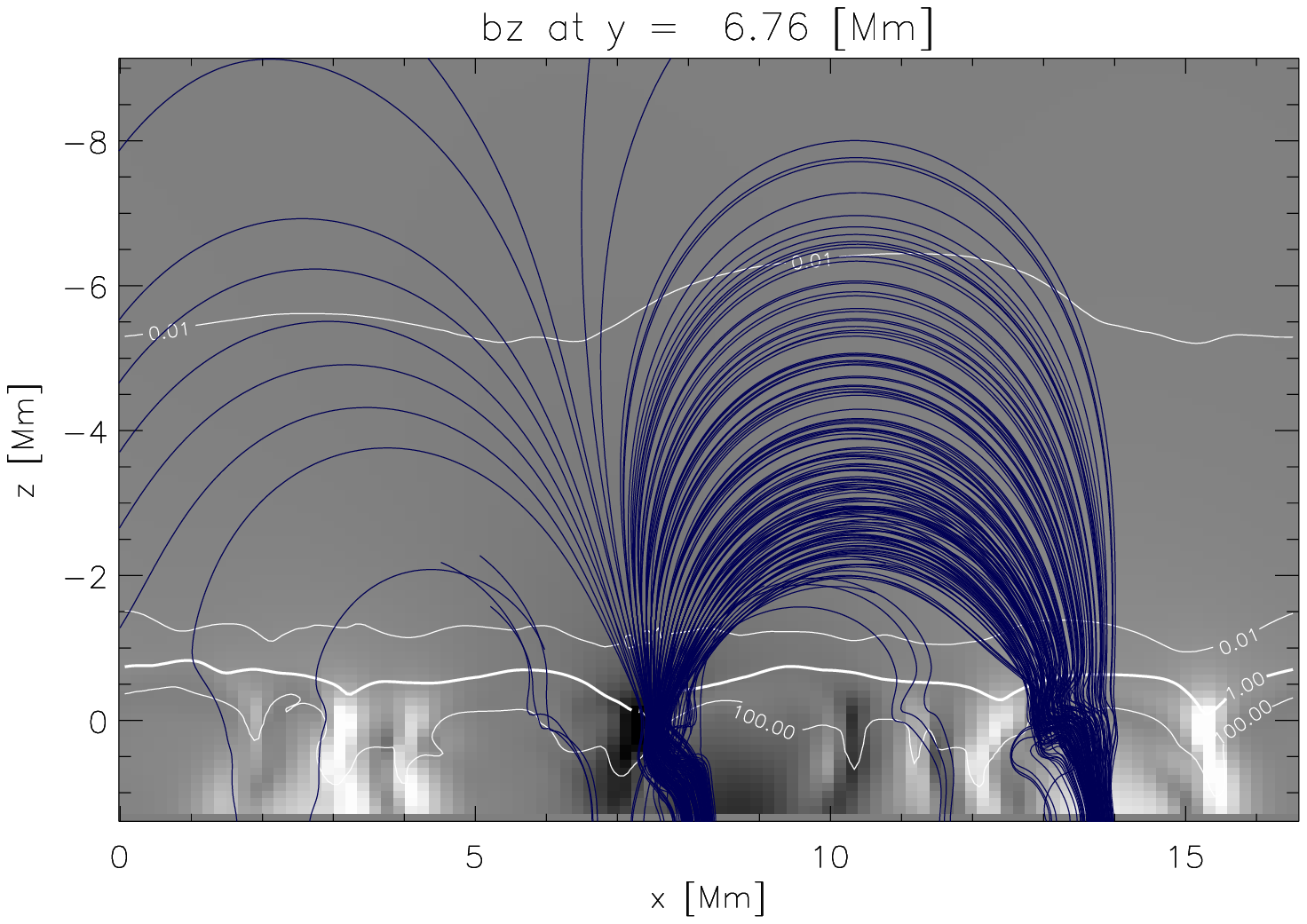}
  \caption[]{\label{hansteen-fig:qsmag-256_xy_xz_vec_104_005}
  The ${\mathbf B}_z$ component of the magnetic field
  in the photosphere, $z=0$~Mm is shown in the upper panel. A subset of magnetic 
  field lines -- a semi-random selection based on the strongest magnetic field at the
  height where $\beta\approx 1$ is drawn. The lower panel shows the same, 
  projected from side. In addition countours of constant $\beta$ are drawn in white.
  Note that the field lines identified
  fit well with the location of maximum emission in the Ne~{\sc viii} and Mg~{\sc X}
  lines as evident in Figure~\ref{hansteen-fig:o6_ne8_mg10}. 
}\end{figure}

\section{Method}

There are several reasons that the attempt to construct forward models
of the convection zone or photosphere to corona  system has been so long
in coming. We will mention only a few:

The magnetic field
will tend to reach heights of approximately the same as the distance
between the sources of the field. Thus if one wishes to model the corona
to a height of, say, 10~Mm this requires a horizontal size close to the
double, or 20~Mm in order to form closed field regions up to the upper
boundary. On the other hand, resolving photospheric scale heights of
100~km or smaller and transition region scales of some few tens of
kilometers will require minimum grid sizes of less than 50~km,
preferably smaller. (Numerical ``tricks'' can perhaps ease some of
this difficulty, but will not help by much more than a factor two).
Putting these requirements together means that it is difficult to get
away with computational domains of much less than 150$^3$ --- a
non-trivial exercise even on todays systems.

The ``Courant condition'' for a
diffusive operator such as that describing thermal conduction scales
with the grid size $\Delta z^2$ instead of with $\Delta z$ for the
magneto-hydrodynamic operator. This severely limits the time step $\Delta
t$ the code can be stably run at. One solution is to vary the magnitude
of the coefficient of thermal conduction when needed. Another, used in 
this work, is to
proceed by operator splitting, such that the operator advancing the
variables in time is $L=L_{\rm hydro}+L_{\rm conduction}$, then solving 
the conduction operator implicitly, for example using the multigrid method. 

Radiative losses from the
photosphere and chromosphere are optically thick and 
require the solution of the transport equation. A sophisticated
treatment of this difficult problem was devised by 
\citep{Nordlund1982} in which opacities are binned according to their
magnitude; in effect one is constructing wavelength bins that represent
stronger and  weaker lines and the continuum so that radiation in all
atmospheric regions is treated to a certain approximation. 
%
%
If one further assumes that opacities are in LTE
the radiation from the photosphere can be modeled. Modeling the chromosphere 
requires that the scattering of photons is treated with greater care
(\cite{Skartlien2000}), or in addition that one uses methods assuming
that chromospheric radiation can be tabulated as a function of local
thermodynamic variables {\it a priori}.

\begin{figure}  
  \centering
  \includegraphics[width=\textwidth]{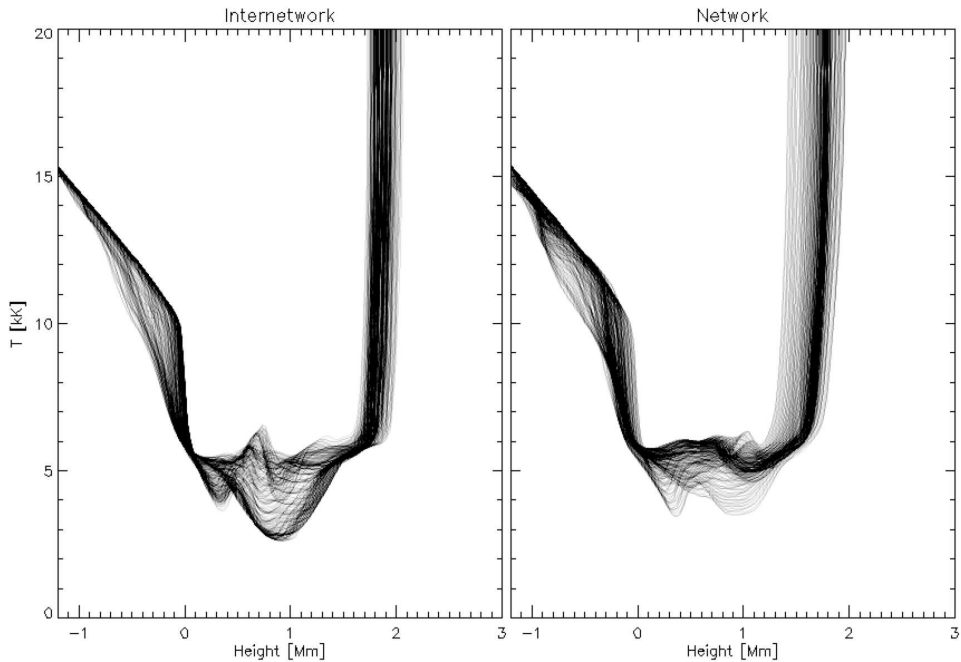}
  \caption[]{\label{hansteen-fig:tdist-chrom}
  Temperature structure in the chromosphere as found above 
  a region of very weak photospheric field ({\it left panel}) and above a region of 
  stronger magnetic field ({\it right panel}). Note that the average chromospheric temperature is higher
  and that the transition region extends further down towards the photosphere in the `network'.
  Note also the evidence of shock propagation in the left `internetwork' panel.
}\end{figure}

\begin{figure}  
  \centering
  \includegraphics[width=\textwidth]{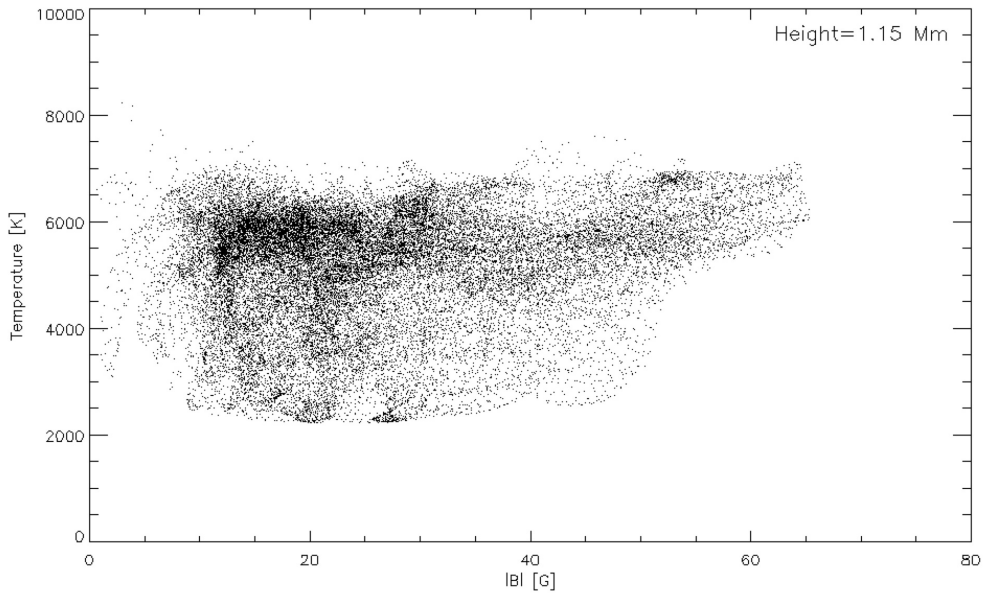}
  \caption[]{\label{hansteen-fig:B-Tg-1.15Mm-1}
  Relation between the magnetic field strength and temperature in the chromosphere
  1.15~Mm above the photosphere.
}\end{figure}

In this paper we have used the methods mentioned to solve the MHD equations, 
including thermal conduction and non-grey non-LTE radiative transfer. 
The numerical scheme used is an extended version of the numerical code described
in \cite{Dorch+Nordlund1998,Mackay+Galsgaard2001} and in more detail
by Nordlund \& Galsgaard at http://www.astro.ku.dk/$\sim$kg. 
In short, the code functions as follows: The variables are represented
on staggered meshes, such that the density $\rho$ and the internal energy
$e$ are volume centered, the magnetic field components ${\bf B}$ and the momentum densities
$\rho{\bf u}$ are face centered, while the electric field ${\bf E}$
and the current ${\bf J}$ are edge centered. A sixth order accurate 
method involving the three nearest neighbor points on each side is
used for determining the partial spatial derivatives. In the cases where
variables are needed at positions other than their defined positions 
a fifth order interpolation scheme is used. The equations are stepped 
forward in time using the explicit 3rd order predictor-corrector 
procedure by \cite{Hyman1979}, modified for variable time steps. 
In order to suppress numerical noise, high-order artificial diffusion is added both
in the forms of a viscosity and in the form of a magnetic diffusivity.

\section{3d Models}


The models described here are run on a box of dimension 16$\times$8$\times$16~Mm$^3$ 
resolved on a grid of $256\times 128\times 160$ points, equidistant in $x$ and $y$ but with
increasing grid size with height in the $z$ direction. At this resolution the model 
has been run a few minutes solar time starting from an earlier simulation with half the 
resolution presented here. The lower resolution simulation had run some 20~minutes solar 
time, starting from a (partially) relaxed convective atmosphere in which a potential 
field with field strengths of order 1~kG at the lower
boundary and an average unsigned field strength of 100~G in the photosphere 
was added. The convective atmosphere has been built up from successively 
larger models, and has run of order an hour solar time; some periodicities are still 
apparent at lower heights where the time scales are longer (of order several hours near
the lower boundary).

The initial potential magnetic field was designed to have
properties similar to those observed in the solar photosphere. The average
temperature at the bottom boundary is maintained by setting the entropy
of the fluid entering through the bottom boundary. The bottom boundary,
based on characteristic extrapolation, is otherwise open, allowing fluid
to enter and leave the computational domain as required. The magnetic
field at the lower boundary is advected with the fluid. As the
simulation progresses the field is advected with the fluid flow in the
convection zone and photosphere and individual field lines quickly
attain quite complex paths throughout the model. 

A vertical cut of the temperature structure in the model is shown in
Figure~\ref{hansteen-fig:qsmag-256t_xz_tg}. 
In Figure~\ref{hansteen-fig:fig_icont} we show the emergent intensity in various continua
as calculated {\it a posteriori} from a data cube some minutes into the simulation 
run. Though the analysis of these intensities is far from complete (the model is 
still in some need of further relaxation) a number of observed solar characteristics
are recognized. Solar granulation seems faithfully reproduced in the 300~nm and 500~nm
bands including bright patches/points in intergranular lanes where the magnetic field is
strong. Reverse granulation is evident in the 170~nm band as is enhanced
emission where the magnetic field is strong. Bright emission in the bands formed higher
in the chromosphere is a result of both strong magnetic fields as well as hydrodynamic 
shocks propagating through the chromosphere.

The field in the models described here was originally potential. However, it is rapidly 
deformed by convective motions in the photosphere and below and becomes concentrated in
down-flowing granulation plumes on a granulation time-scale. In regions below 
where $\beta\equiv{p_g/p_B}=1$ the magnetic field is at the mercy of plasma motions, 
above the field expands, attempts to fill all space, and forms loop like structures. 
The $B_z$ component of the magnetic field in the photosphere 
is shown in the upper panel of Figure~\ref{hansteen-fig:qsmag-256_xy_xz_vec_104_005}, also 
plotted are magnetic field lines chosen on the basis of their strength at the surface where 
$\beta=1$. The same field lines seen from the side are plotted in the lower panel of 
Figure~\ref{hansteen-fig:qsmag-256_xy_xz_vec_104_005} overplotted the vertical magnetic field $B_z$.

\begin{figure}  
  \centering
  \includegraphics[width=\textwidth]{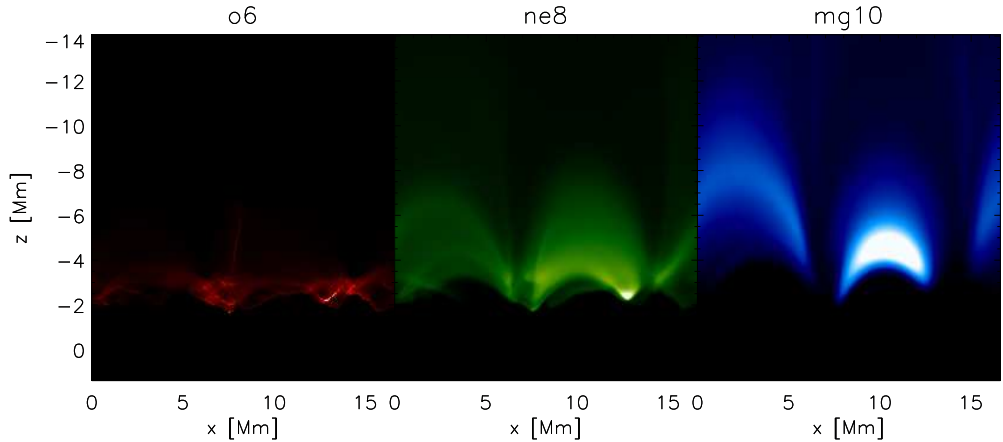}
  \caption[]{\label{hansteen-fig:o6_ne8_mg10}
    Line emission from the O~{\sc vi}, 300~kK formation temperature, 
    Ne~{\sc viii} line, formed at roughly 700~kK, and Mg~{\sc x} formed at some 
    1~MKk, all seen seen from the side. Note the ray of O~{\sc vi} emission near
    $x=7$~Mm that extends to some $z=-6$~Mm.
}\end{figure}

Chromospheric energetics and dynamics are set by a number of factors. Among 
the most important of these are acoustic waves generated in the photosphere and 
convection zone impinging on the chromosphere from below; the topology of the magnetic
field and the location of the plasma $\beta=1$ surface; the amount of chromospheric
heating due the dissipation of magnetic energy; non-LTE radiative losses and related
phenomena such as time dependent ionization and recombination. Most of these phenomena 
with the exception of time dependent ionization is accounted for (to various degrees
of accuracy) in the models presented here. The latter is currently under 
implementation (\cite{Leenaarts+etal2007}). Examples of the chromospheric temperature 
structure and its relation to the magnetic field are shown in Figures~\ref{hansteen-fig:tdist-chrom} 
and \ref{hansteen-fig:B-Tg-1.15Mm-1}.

As the stresses in the coronal field grow so does the energy density of 
the field. This energy must eventually be dissipated; at a rate commensurate with 
the rate at which energy flux is pumped in. This will depend on the strength of the magnetic
field and on the amplitude of convective forcing. On the Sun the magnetic diffusivity 
$\eta$ is very small and gradients must become very large before dissipation 
occurs; in the models presented here we operate with an $\eta$ many orders 
of magnitude larger than on the Sun and dissipation starts at much smaller magnetic
field gradients. Even so, it seems the model is able to reproduce diagnostics that 
resemble those observed in the solar transition region and corona as shown in 
Figure~\ref{hansteen-fig:o6_ne8_mg10}. (It is also interesting to note that we find emission 
from O~{\sc vi} 103.7~nm in a narrow ray up to 6~Mm above the photosphere, much higher
than it should be found in a hydrostatically stratified model.)

\section{Conclusions}

The model presented here seems a very promising starting point and tool for achieving
an understanding of the outer solar layers. But perhaps a word or two of caution 
is in order before we celebrate our successes. Are the tests we 
are subjecting the model to --- {\it e.g.} the comparison of synthetic observations 
with actual observations actually capable of separating a correct description of
the sun from an incorrect one? Conduction along field lines will naturally make loop like 
structures. This implies that reproducing TRACE-like ``images'' is perhaps not
so difficult after all, and possible for a wide spectrum of coronal heating 
models. The transition region diagnostics are a more discerning test, but clearly
it is still too early to say that the only possible coronal model has been 
identified. It will be very interesting to see how these forward coronal heating
models stand up in the face of questions such as: How does the corona react to 
variations in the total field strength, or the total field topology, and what 
observable diagnostic signatures do these variations cause? One could also wonder 
about the role of emerging flux in coronal heating: How much new magnetic flux 
must be brought up from below in order to replenish the dissipation of field 
heating the chromosphere and corona?

\acknowledgements 
This work was supported by the Research Council of Norway grant
146467/420 and a grant of computing time from the Program for
Supercomputing. 

\bibliographystyle{cspm-bib}      
\bibliography{coimbra_pg.bib}  

\end{document}